# Ion Polarization Scheme for MEIC

A.M. Kondratenko[1], M.A. Kondratenko[1], Yu.N. Filatov[2],
Ya.S. Derbenev[3], F. Lin[3], V.S. Morozov[3], and Y. Zhang[3]

[1]*Science and Technique Laboratory "Zaryad", Novosibirsk 630090, Russia*
[2]*Moscow Institute of Physics and Technology, Dolgoprudny 141700, Russia*
[3]*Thomas Jefferson National Accelerator Facility, Newport News, VA 23606, USA*
Dated: June 12, 2015

## CONTENTS

# 1. INTRODUCTION

The design requirements for the ion beam polarization in MEIC are [1, 2]:
- High polarization (over 70%) for protons or light ions (d, $^3$He$^{++}$, and possibly $^6$Li$^{+++}$)
- Both longitudinal and transverse polarization at all interaction points
- Sufficiently long polarization lifetime
- Spin flipping at a high frequency

The choice of a figure 8 shape for the booster and collider rings of MEIC opens wide possibilities for preservation of the ion polarization during beam acceleration as well as for control of the polarization at the collider's interaction points [3-8]. As in the case of accelerators with Siberian snakes, the spin tune is energy independent but is equal to zero instead of one half. The figure-8 topology eliminates the effect of arcs on the spin motion. There appears a unique opportunity to control the polarization of any particle species including deuterons, using longitudinal fields of small integrated strength (weak solenoids). Contrary to existing schemes, using weak solenoids in figure-8 colliders, one can control the polarization at the interaction points without essentially any effect on the beam's orbital characteristics. A universal scheme for control of the polarization using weak solenoids provides an elegant solution to the problem of ion acceleration completely eliminating resonant beam depolarization. It allows one to easily adjust the polarization in any direction at any orbital location, which becomes necessary when transferring the beam from one ring into another or when measuring the polarization by polarimeters. It also allows for an easy manipulation of the spin direction at an interaction point during an experiment. The latter feature allows one to set up a spin-flipping system with a spin reversal time of less than a second. By compensating the coherent part of the zero-integer spin resonance strength, which arises due to errors in alignment of the magnetic element of the lattice, one can reduce the field integrals of the control solenoids by a few orders of magnitude.

The above advantages of a figure-8 collider with weak spin control solenoids allow polarized beam studies in MEIC at an unprecedented precision level.

Below we will first discuss the main characteristic features of the spin behavior in figure-8 colliders and will then describe the baseline for ion polarization control in the accelerator complex of MEIC.

# 2. COLLIDERS WITH "PREFERRED SPIN AXIS" AND "TRANSPARENT TO THE SPIN"

Describing the polarization dynamics of ion beams in colliders (cyclic accelerators, storage rings) is in many respects similar to describing their orbital motion. An analog of the closed orbit, along which the particle velocity vector is periodic, is the periodic spin precession axis $\vec{n}$. When a particle is moving along the closed orbit, its spin initially directed along the $\vec{n}$ axis will restore its orientation every particle turn. Thus, the effect of magnetic fields along the closed orbit on the spin represents a rotation about the $\vec{n}$ axis by some angle $2\pi\nu$, which determines the spin tune $\nu$. When the particle trajectory deviates from the closed orbit, the spin precession axis as well as the spin tune also experience deviations. The beam polarization is stable just along the collider's $\vec{n}$ axis. The particle spins transverse to the $\vec{n}$ axis get completely disordered in a few hundred particle turns due to the spin tune spread leading to beam depolarization.

From the spin dynamics point of view, all colliders can be generally categorized into two types, namely, colliders "with preferred spin direction" and colliders "transparent to the spin".

In colliders "with preferred spin direction", the periodic spin motion along the closed orbit is unique, i.e. the static magnetic lattice determines a single stable orientation of the beam polarization. The fractional part of the spin tune differs from zero.

Let us give examples of colliders (accelerators) with preferred spin direction.

1. *Conventional accelerator.* Conventional accelerators include accelerators consisting of arcs and straight sections. The stable polarization direction is vertical while the spin tune is proportional to the beam energy, which unavoidably leads to crossing of many spin resonances during acceleration and, consequently, to resonant depolarization of the beam.

2. *Collider with a Siberian snake.* Introduction of Siberian snakes opens wide opportunities for obtaining of intense polarized beams at high energies. Insertion of one Siberian snake into a collider ring completely rearranges the spin dynamics. Instead of being vertical, polarization becomes stable in the collider's plane while the spin tune does not depend on the energy at all and is equal to one half. A Siberian snake eliminates crossing of spin resonances and completely solves the problem of polarization preservation during beam acceleration [9-10].

3. *Collider with two Siberian snakes.* Insertion of two Siberian snakes into opposite straights of a collider allows one to keep the polarization vertical in the arcs. The spin tune is then also independent of energy, is determined by the angle between the snakes' spin rotation axes, and can vary in the range from zero to one half. In a conventional scheme with two Siberian snakes, the angle between the snakes' rotation axes is equal to 90° and the spin tune is equal to one half, the same as in the case of a collider with one snake (RHIC, BNL) [11-13].

In colliders "transparent to the spin", any spin direction repeats every particle turn along the closed orbit, i.e. the accelerator's magnetic lattice is transparent to the spin. The fractional part of the spin tune is equal to zero.

Let us give examples of colliders (accelerators) transparent to the spin.

1. *Conventional accelerator (resonant case).* As noted above, conventional accelerators belong to the type of accelerators "with preferred spin direction" everywhere except narrow energy bands in the regions of integer resonances $\nu = \gamma G = k$, i.e. when the combined effect of arcs on the spin results in an integer number $k$ of rotations about the vertical axis.

2. *Collider with two identical Siberian snakes.* When using two identical snakes with the angle between their rotation axes equal to zero or 180°, the fractional part of the spin tune becomes equal to zero. A project of a collider with two solenoidal snakes has been proposed at Dubna (NICA, JINR) [14].

3. *Collider in the shape of a figure 8.* The most natural representative of a collider "transparent to the spin" is an accelerator in the shape of a figure 8 (MEIC, JLAB) [1]. The resulting effect of the arc magnets on the spin dynamics over a particle turn reduces to zero and the ring becomes "transparent" to the spin: the spin first rotates about the vertical field in the first arc and then its rotation is compensated by an opposite field in the second arc. Any spin direction at any orbital location repeats every turn. The spin tune is zero. The particles are in the region of a zero-integer spin resonance $\nu = 0$.

From the above examples, we see that a collider of one type can be easily transformed into a collider of the other type. For instance, insertion of a Siberian snake into a spin-transparent collider converts it into a collider with preferred spin direction. Inversely, addition of a second identical snake to a single-snake collider with preferred spin direction converts it into a spin-transparent collider.

Colliders transparent to the spin offer a unique opportunity to efficiently control the ion polarization using small magnetic field integrals. In such a collider, any small perturbation has a strong effect on the beam polarization. To stabilize the spin direction, one must introduce additional fields into the collider's lattice, which "shift" the spin tune by a small value ($\nu \ll 1$) and set the necessary orientation of the $\vec{n}$ axis. The required field integrals (weak fields) are significantly lower than the field integrals used in Siberian snakes (strong fields) and are limited by the strength of the zero-integer spin resonance $w_0$: $\nu \gg w_0$.

For example, for a collider transparent to the spin with a maximum energy of 100 GeV, it is sufficient to introduce a solenoid with a field integral of about 10 T·m to stabilize the polarization in the longitudinal direction at the solenoid insertion place. Besides, such a scheme is universal for all particle species including both protons and deuterons. In RHIC, longitudinal proton polarization is obtained using two helical snakes and two spin rotators with a combined transverse field integral of $2 \times 24 + 2 \times 23 \approx 100$ T·m. For deuterons, due to the small value of their anomalous magnetic moment, the total transverse field integral in RHIC would have to be about 2500 T·m. In a collider with one solenoidal snake, obtaining longitudinal polarization of protons and deuterons would require longitudinal field integrals of about 360 and 1000 T·m, respectively. These examples demonstrate that obtaining longitudinal deuteron polarization in colliders with preferred spin direction presents a serious problem.

In colliders, a crucial task is not only preservation of the polarization during the energy ramp but also its control at the interaction points *during experiments*.

Traditionally, in colliders with preferred spin direction, polarization is controlled using a pair of spin rotators with "strong" fields, which provide the required spin orientation at an interaction point and then return the spin to its original direction. This process changes not only the polarization direction but the trajectories of the beam particles as well. This causes a change in the orbital characteristics of the beam. There are betatron tune shifts, changes in dispersion, beta-functions, luminosity, etc. Accounting for the changes in the beam orbital characteristics becomes a problem when doing experiments with polarized beams in colliders.

In colliders with spin transparency, ion polarization can be efficiently controlled using "weak" fields, which have essentially no effect on the beam's orbital characteristics. What especially stands out is the possibility of using weak solenoids, which do not impact the closed orbit at all. In the collider's energy range of up to 100 GeV, the field integrals of these solenoids are approximately two orders of magnitude lower than the field integrals of the spin rotators with strong fields. There is no problem with changing the field of such solenoids during adjustment of the beam polarization direction. It becomes possible to reverse the spin in less than a second that allows for polarized beam experiments at a new precision level.

Let us demonstrate the main advantages of spin transparent figure-8 colliders when solving the problems of polarization preservation and spin manipulation during experiments in the ion complex of MEIC.

## 3. ION POLARIZATION IN THE MEIC ACCELERATOR COMPLEX

MEIC is designed to be a traditional ring-ring collider. The central part of this facility is a set of figure-8 collider rings as shown in Fig. 1 [2]. The electron collider ring is made of normal conducting magnets reconditioned from the decommissioned PEP-II e$^+$e$^-$ collider at SLAC, and will store an electron beam of 3 to 10 GeV. The ion collider ring is made of new super-ferric magnets, a cost-effective type of superconducting magnet with modest field strength (up to 3 T) and will store a beam with energy of 20 to 100 GeV for protons or up to 40 GeV per nucleon for heavy ions. The two collider rings are stacked vertically and housed in the same underground tunnel. They have nearly identical circumferences of approximately 2.2 km. The electron and ion collider rings intersect at two symmetric points, one in each of the two long straights; thus, two detectors can be accommodated. The ion beam executes a vertical excursion to the plane of the electron ring to realize a horizontal crossing for electron-ion collisions. The two long straights also support other utility elements of the collider rings, among them the injection/ejection systems, the RF systems, the electron cooler, and the beam polarimeters.

In the present conceptual design, the MEIC ion complex consists of sources for polarized light ions and non-polarized light to heavy ions, a 280 MeV pulsed SRF ion linac, a 8 GeV kinetic energy booster ring, and a medium-energy collider ring (see Fig. 1). All of the above energy parameters are for the proton beam; they must be scaled appropriately for ion beams using the mass-to-charge ratio for the same magnetic rigidity.

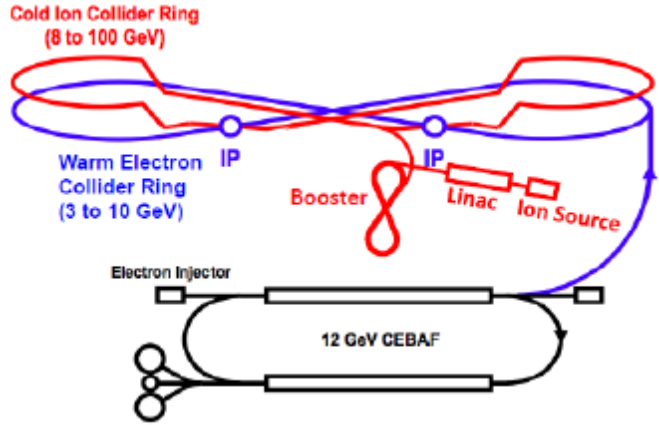

**Figure 1:** The MEIC electron and ion collider rings and the large ion booster ring are stacked vertically and housed in the same tunnel.

For experiments with polarized ion beams in the MEIC complex, one must solve the problem of preserving the polarization at all stages of ion beam acceleration including the sources of polarized light ions, SRF ion linac, booster, and collider ring. One must also solve the problem of ion polarization control at the collider's interaction points.

### 3.1 PRESERVATION OF THE ION POLARIZATION IN THE BOOSTER

In figure-8 accelerators, the spin tune is zero and is independent of energy. Similar to an accelerator with a Siberian snake, a figure-8 accelerator eliminates the possibility itself of crossing spin resonances during an energy ramp. To preserve the beam polarization during acceleration, it is sufficient to stabilize the spin motion in the zero-integer spin resonance region using one weak solenoid. Figure 8 shows schematics of polarization preservation in the booster of the MEIC complex [5]. A solenoid stabilizes the longitudinal polarization direction in the straight where it is installed. Injection and extraction of the beam take place in the same accelerator straight. Polarization must then be matched to the longitudinal direction at injection of the ion beam from the linac into the booster and at extraction from the booster into the collider ring.

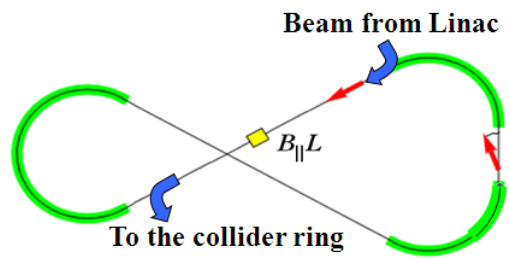

**Figure 2:** Acceleration and spin matching in the booster.

The spin tune value set by the weak solenoid must greatly exceed the strength of the zero-integer spin resonance. To stabilize the spin and orbital motions, the solenoid field must change proportionally to the beam momentum. Table 1 shows parameters of the solenoid providing preservation of the proton and deuteron polarizations in the whole energy ranges of the booster.

**Table 1.** Parameters of the solenoid preserving the polarization in the booster.

| $p_{inj}/p_{ext}$, GeV/c | $(B_\parallel L_\parallel)_{inj}/(B_\parallel L_\parallel)_{ext}$, T·m | $L_\parallel$, cm | $\nu_{deut}/\nu_{prot}$ |
|---|---|---|---|
| 0.785/9 | 0.06/0.7 | 60 | 0.003/0.01 |

The required solenoid field integral does not exceed 1 T·m at the top energy of the booster. There is no problem with ramping up the fields of such solenoids during the acceleration cycle. To avoid crossing of spin resonances up to the momentum of 9 GeV/c, accelerators with Siberian snakes require a solenoidal snake with a field integral of about 33 T·m for protons and about 110 T·m for deuterons. Note the universality of the polarization preservation scheme with regard to the particle species. Figure-8 colliders allow one to easily preserve the deuteron polarization up to the maximum energy of the complex.

As pointed out above, weak solenoids essentially do not change the orbital parameters of the ring. As an example, Fig. 3 shows graphs of the $\beta$-functions and vertical dispersion $D_y$ in the booster for the cases without (top) and with (bottom) insertion of the solenoid. The insertion location is circled in red.

The radial and vertical betatron tunes of the booster without the solenoid are $\nu_x = 7.9774$ and $\nu_y = 6.7928$, respectively. The solenoid causes betatron tune shifts, which are independent of energy. In the approximation of a short solenoid ($L_\parallel \ll \beta$), the shifts of the radial and vertical betatron tunes can be estimated using the formula

$$\Delta \nu_{x,y} = \frac{1}{16\pi} \frac{(B_\parallel L_\parallel)^2}{(B\rho)^2} \frac{\beta_{x,y}}{L_\parallel} = const.$$

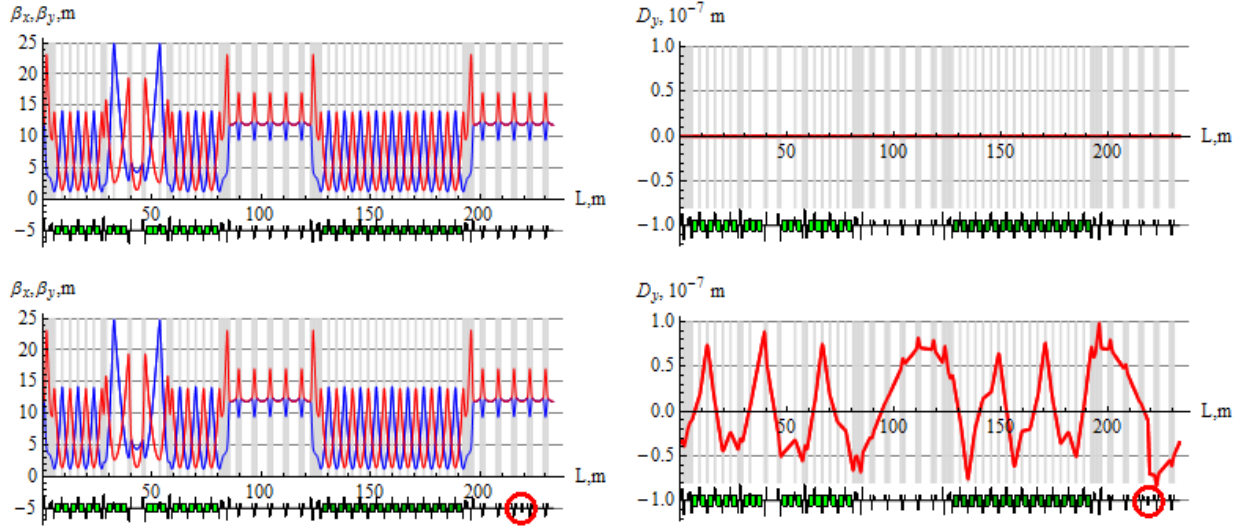

**Figure 3:** $\beta$-functions and vertical dispersion $D_y$ for the cases without (top) and with (bottom) insertion of the solenoid in the booster.

For the maximum solenoid field integral of 0.7 T·m at the booster's maximum rigidity of $B\rho = 30$ T·m and for $\beta_{x,y} = 12$ m, the betatron tune shift is about $\Delta\nu_{x,y} \approx 2 \times 10^{-4}$.

The considered example shows that the influence of a weak solenoid on the orbital parameters of the booster is negligibly small despite even the fact that one of the betatron tunes is near an integer, i.e. when parameters of the orbital motion are particularly sensitive to small orbit perturbations.

### 3.2. ION POLARIZATION CONTROL IN THE COLLIDER RING

The advantages of using weak solenoids in figure-8 colliders stand out most prominently when manipulating the particle spin at the interaction points during an experiment [4].

As already noted, in a collider transparent to the spin, the spin tune and stable polarization direction are determined not by the structural arc magnets but by small solenoids introduced into the collider's lattice. The small solenoids do not change the design orbit and allow for control of the beam polarization without essentially any effect on the parameters of the orbital motion.

Weak solenoid and dipole inserts can be used to construct a universal 3D spin rotator, which allows for adjustment of any polarization direction of any particle species (p, d, $^3$He, ...) in the collider.

Figure 4 shows a schematic of the 3D spin rotator [5] for ion polarization control located at the end of the experimental straight. The rotator consists of three modules: those for control of the radial $n_x$, vertical $n_y$, and longitudinal $n_z$ components of the polarization (see Fig. 4a). The 3D spin rotator placement in the MEIC ring is shown schematically in Fig. 4b.

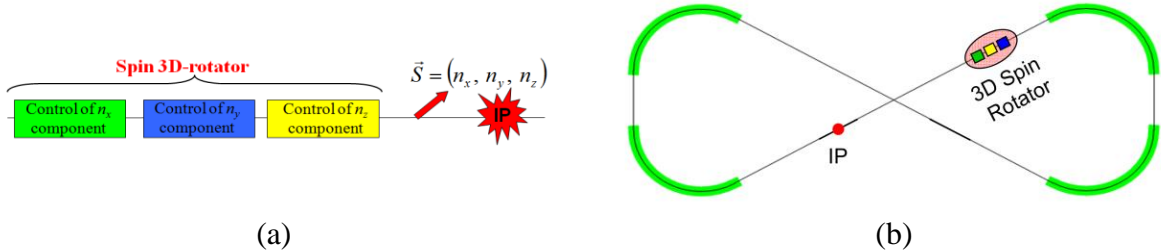

(a)  (b)

**Figure 4:** (a) 3D spin rotator schematic. (b) Spin rotator placement in the ion collider ring.

Figure 5a shows the module for control of the radial polarization component $n_x$, which consists of two pairs of opposite-field solenoids and three vertical-field dipoles producing a fixed orbit bump. The control module for the vertical polarization component $n_y$ is the same as that for the radial component except that the vertical-field dipoles are replaced with radial-field ones (Fig. 5b). To keep the orbit bumps fixed, the fields of the vertical- and radial-field dipoles must be ramped proportionally to the beam momentum. The module for control of the longitudinal polarization component $n_z$ consists of a single weak solenoid (Fig. 5c). There is a substantial flexibility in the placement and arrangement of these modules in the collider. For instance, to free up the space in the experimental straight, the module for control of the vertical polarization component can be installed anywhere in the arc.

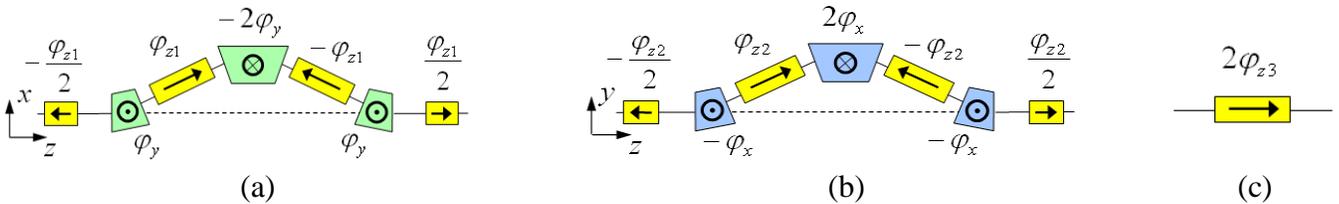

(a)  (b)  (c)

**Figure 5:** Modules for control of the radial (a), vertical (b), and longitudinal (c) spin components.

Under the approximation of a small spin tune, the required spin rotation angles of the solenoids $\varphi_{zi}$ are determined by the following equations ($|\vec{n}| = 1$):

$$\varphi_{z1} = \pi\nu \frac{n_x}{\sin\varphi_y}, \quad \varphi_{z2} = \pi\nu \frac{n_y}{\sin\varphi_x}, \quad \varphi_{z3} = \pi\nu\, n_z.$$

where $n_x$, $n_y$, and $n_z$ are the radial, vertical, and longitudinal polarization components, respectively, at the rotator's exit, $\varphi_x = \gamma G \alpha_x$ and $\varphi_y = \gamma G \alpha_y$ are the spin rotation angles of the aforementioned radial- and vertical-field dipoles, respectively, and $\alpha_x$ and $\alpha_y$ are the respective orbit bending angles of these dipoles. The calculation assumes that, with the solenoids off, the spin tune in the collider is zero.

Schematic placement of the 3D rotator elements in the collider ring's experimental straight is shown in Fig. 6. The lattice quadrupoles are shown in black, the vertical-field dipoles are green, the radial-field dipoles are blue, and the control solenoids are yellow. With each module's length of ~8 m ($L_x = L_y = 0.6$ m, $L_z = 2$ m), the fixed orbit deviation in the bumps is ~16 mm in the whole momentum range of the collider. Placement of each bump between lattice quadrupoles keeps the experimental straight dispersion-free.

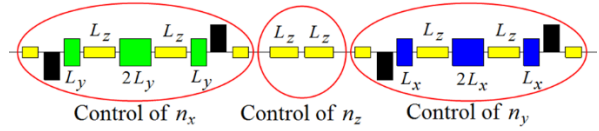

**Figure 6**: Placement of the 3D spin rotator elements.

The maximum required dipole and solenoid magnetic field strengths are 3 and 2 T, respectively. The spin rotator shifts the proton and deuteron spin tunes from zero by sufficient amounts of 0.01 and $10^{-4}$, respectively.

Figure 7 shows dependencies of the solenoid fields in the 3D rotator on the deuteron and proton beam momenta for radial, vertical and longitudinal polarizations in MEIC.

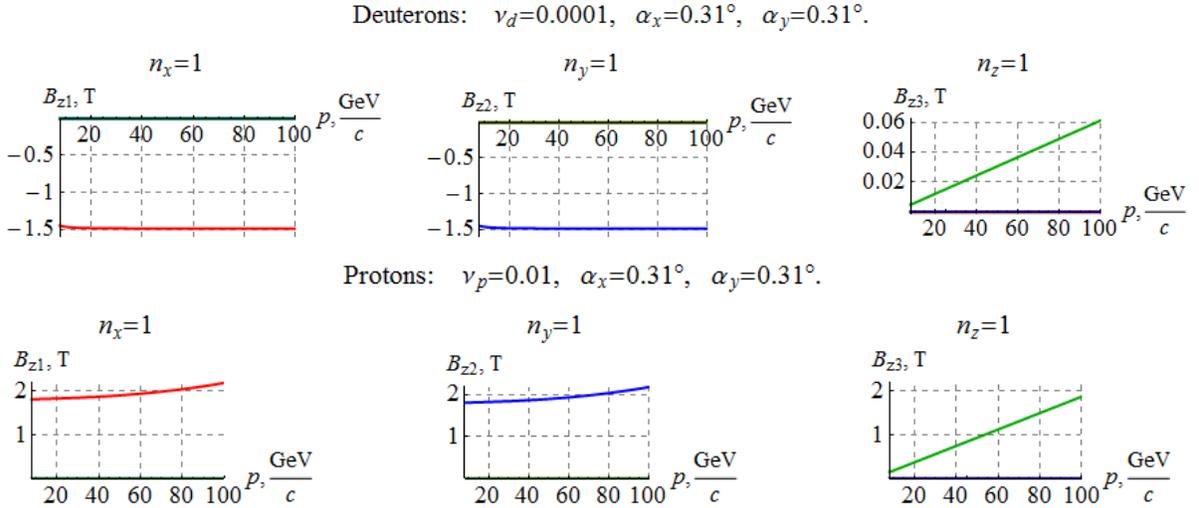

**Figure 7:** Dependencies of the solenoid fields on the deuteron and proton beam momentum in MEIC for the cases of radial, vertical and longitudinal polarizations at the exit from the 3D rotator.

Note that, for the transverse polarization directions, the solenoid fields remain constant at about -1.5 T·m in the whole momentum range of the deuteron beam. For the longitudinal polarization, the solenoid field is proportional to the deuteron momentum and is about an order of magnitude lower than the solenoid fields required for the transverse polarization.

The 3D rotators are designed using weak solenoids and allow performance of the following tasks: matching of the polarization direction at injection, polarization preservation during acceleration and storage, measurement of the beam polarization at any orbital location, and spin manipulation at the interaction point during experimental running.

### 3.3. SPIN FLIPPING IN THE ION COLLIDER RING

The universal 3D spin rotator can be used to arrange a spin-flipping system in MEIC, which provides multiple beam polarization reversals at the interaction point during an experiment [7].

In setting up multiple polarization reversals, it is important to keep the spin tune constant during the reversals. This allows one to avoid multiple crossings of higher-order resonances and thus eliminate resonant beam depolarization.

Consider, for example, polarization reversals in the vertical plane of the collider when polarization components at the interaction point are given by the angle $\Psi$ between the velocity direction and the beam polarization:

$$n_x = 0, \ n_y = \sin \Psi, \ n_z = \cos \Psi.$$

In this case, polarization is controlled by adjusting the fields $B_{z2}$ and $B_{z3}$ in the $n_y$ и $n_z$ polarization control modules. In a $(B_{z2}, B_{z3})$ field diagram, points of constant spin tune represent an ellipse.

Figure 8 shows such an ellipse for protons and deuterons at 100 GeV/c. The calculation assumes that there are no additional dipoles and solenoids between the 3D spin rotator and interaction point and that the spin tune is $\nu = 10^{-2}$ for protons and $\nu = 10^{-4}$ for deuterons.

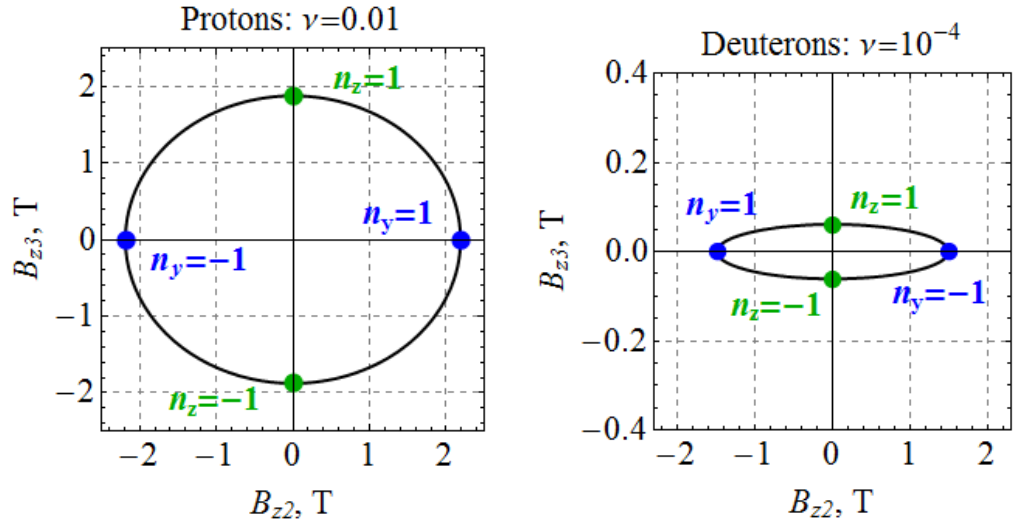

**Figure 8:** Relation of the solenoid fields in the $n_y$ and $n_z$ modules when changing the polarization direction ($\nu$ = const). Shown on the left is the ellipse for the proton beam, on the right is that for the deuteron one. The blue dot marks the solenoid fields, for which $n_y = \pm 1$. The green dot marks those, for which $n_z = \pm 1$.

The dots in Fig. 8 indicate the field values corresponding to the longitudinal ($n_z = \pm 1$, green dots) and vertical ($n_y = \pm 1$, blue dots) polarizations. The field values at these points are listed in Table 2. Thus, by changing the solenoid fields $B_{z2}$ and $B_{z3}$ along the ellipse, we reverse the beam polarization passing sequentially its vertical and longitudinal directions.

**Table 2**. Solenoid fields $B_{z2}$ and $B_{z3}$ providing vertical and longitudinal polarizations of the proton and deuteron beams in the MEIC ion collider ring.

| Polarization | Protons | | Deuterons | |
|---|---|---|---|---|
| | $B_{z2}$, T | $B_{z3}$, T | $B_{z2}$, T | $B_{z3}$, T |
| Vertical $n_y = 1$ | 2.19 | 0 | -1.49 | 0 |
| Vertical $n_y = -1$ | -2.19 | 0 | 1.49 | 0 |
| Longitudinal $n_z = 1$ | 0 | 1.88 | 0 | 0.06 |
| Longitudinal $n_z = -1$ | 0 | -1.88 | 0 | 0.06 |

To preserve the polarization, one then only has to satisfy the condition of adiabatic change of the polarization direction $\vec{n}(B_{zi}) = (n_x, n_y, n_z)$ by the control solenoids $B_{zi}$:

$$\frac{d\vec{n}}{dt} \ll \Omega_{\text{spin}}, \quad \Omega_{\text{spin}} = \nu\, \Omega_c = \text{const},$$

where $\Omega_c$ is the particle revolution frequency in the collider. This condition means that the characteristic spin reversal time in the given examples is limited by 0.1 ms for protons and 10 ms for deuterons. Thus, using solenoids with a field ramp rate of 2 T/s, polarization can be flipped in a second [15].

### 3.4. EFFECT OF 3D SPIN ROTATOR ON THE ORBITAL BEAM PARAMETERS IN MEIC

Effect of the 3D spin rotator is calculated for multiple reversals of the beam polarization in the vertical plane (*yz*) of the detector during an experiment (spin flipping) [16]. Figures 9 and 10 show graphs of the solenoid fields in the $n_y$ and $n_z$ modules of the 3D rotator versus the angle Ψ between the spin and the beam direction for deuterons and protons, respectively.

With such a synchronous change of the solenoid fields, the spin tune remains constant while the polarization direction changes in the (*y*,*z*) plane and is given by the angle Ψ:

$$n_x = 0, \quad n_y = \sin\Psi, \quad n_z = \cos\Psi.$$

The stability of the reversals is provided by keeping the spin tune fixed while changing the spin direction, which eliminates the possibility of crossing spin resonances.

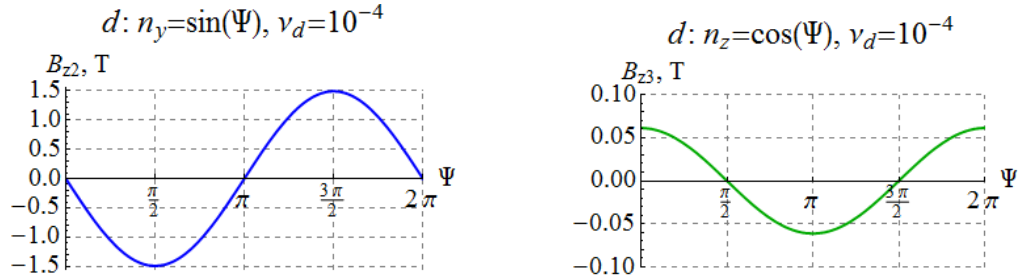

**Figure 9:** : Solenoid fields in the $n_y$ and $n_z$ modules versus the polarization angle Ψ for the deuteron beam.

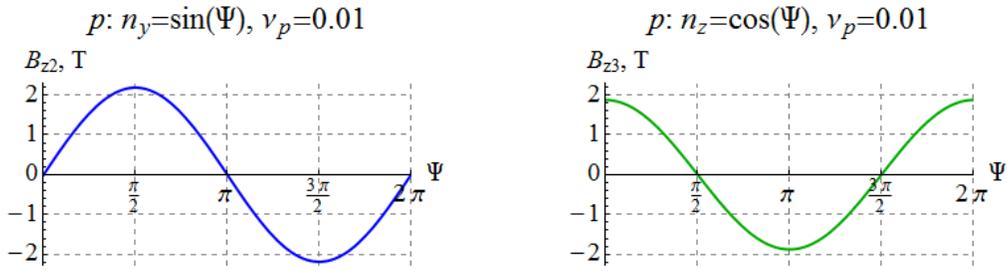

**Figure 10:** Solenoid fields in the $n_y$ and $n_z$ modules versus the polarization angle $\Psi$ for the proton beam.

Figure 11 shows a part of the collider's experimental straight with the 3D spin rotator and interaction point (IP) locations indicated. The figure shows graphs of the horizontal and vertical $\beta$-functions. With the 3D rotator off, the betatron tunes and the $\beta$-function values at the IP are:

$$\nu_x = 24.38, \quad \nu_y = 24.28, \quad \beta_x = 10 \text{ cm}, \quad \beta_y = 2 \text{ cm}.$$

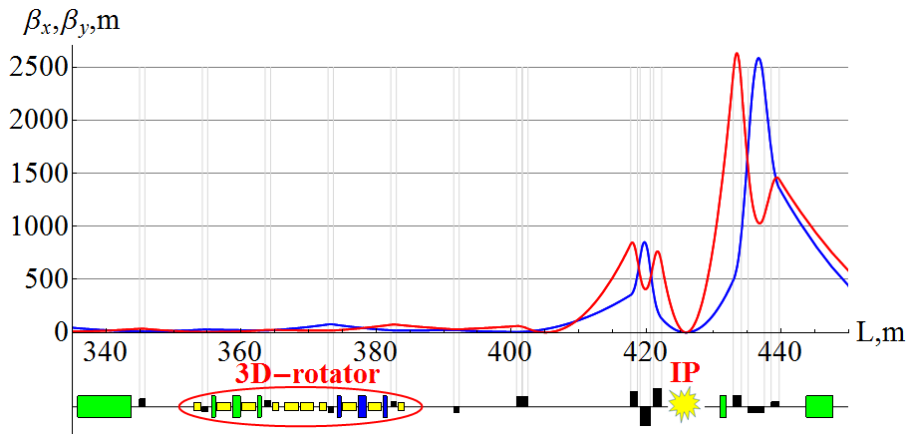

**Figure 11:** $\beta$-functions and 3D rotator placement in the experimental straight of MEIC.

Figure 12 shows change in the $\beta$ functions at the IP for deuterons and protons when changing the spin direction in the vertical plane of the detector during an experiment. As one can see, the maximum change of the $\beta$ functions does not exceed 60 and 200 μm for deuterons and protons, respectively, i.e. the beam size remains virtually the same. Similarly, Fig. 13 shows change in the betatron tunes. One can see that the betatron tune shifts at 100 GeV/$c$ do not exceed $10^{-4}$ for deuterons and $2\times10^{-4}$ for protons.

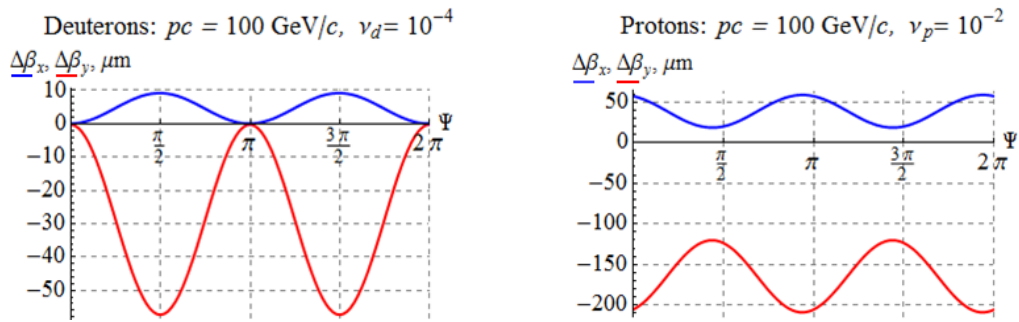

**Figure 12:** Change in the $\beta$ functions at the IP versus the polarization angle.

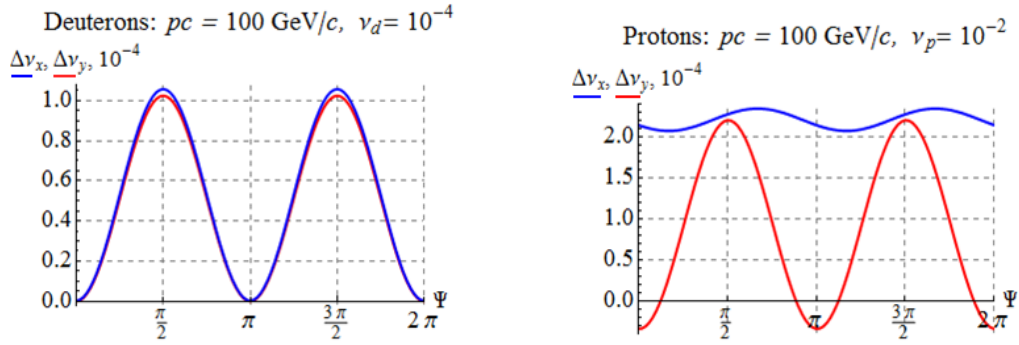

**Figure 13:** Change in the betatron tunes versus the polarization angle.

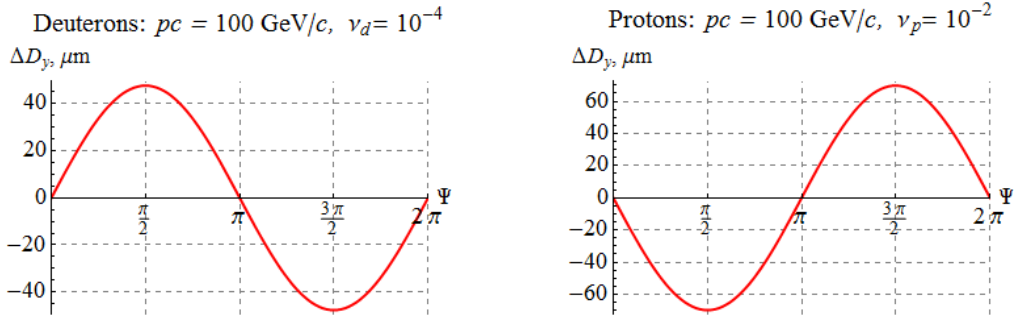

**Figure 14:** Change in the dispersion function at the IP versus the polarization angle for protons and deuterons.

Calculations show that change in the dispersion due to the 3D rotator is also negligibly small. Figure 14 shows change in the dispersion function at the IP when changing the spin direction in the vertical plane of the detector for deuterons and protons. The control solenoids induce vertical dispersion in the collider ring, which, at 100 GeV/$c$, does not exceed 50 and 70 $\mu$m for deuterons and protons, respectively.

Our numerical calculations confirm that the 3D spin rotator does not affect the orbital beam parameters of the MEIC ion collider ring.

### 3.5. CALCULATION OF THE BEAM POLARIZATION IN MEIC

Let us present calculations of the proton and deuteron beam polarizations in the MEIC ion collider ring with a single 3D rotator determining the equilibrium polarization at the interaction point [16].

As an example, in Fig. 15, for an ideal collider structure, the equilibrium polarization components of a 100 GeV/$c$ deuteron beam are shown as functions of the orbital length $z$ around the ring for the case of longitudinal ($n_z(z_{IP}) = 1$) polarization at the interaction point. The blue, red, and green curves show the radial, longitudinal, and vertical polarization components, respectively. Note that the vertical polarization component is zero around the whole ring.

In Fig. 16, for an ideal collider structure, the equilibrium polarization components of a 100 GeV/$c$ proton beam are shown as functions of the orbital length $z$ along the experimental straight for the case of radial ($n_x(z_{IP}) = 1$) polarization at the interaction point. In contrast to the deuteron beam, the radial and longitudinal components of the proton polarization change significantly at each bending magnet of the lattice. The horizontal polarization component undergoes about 127 turns in each arc and is rotated significantly by the vertical-field dipoles located near the interaction point.

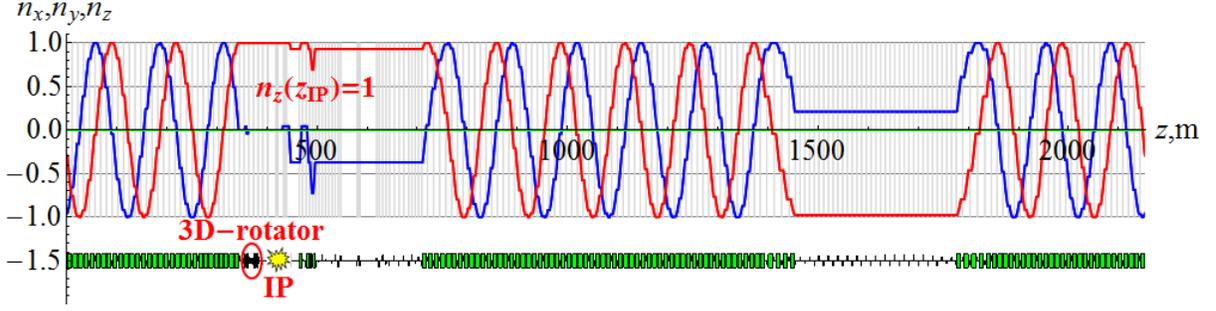

**Figure 15:** Deuteron beam's polarization around the ion collider ring with $n_z(z_{IP}) = 1$.

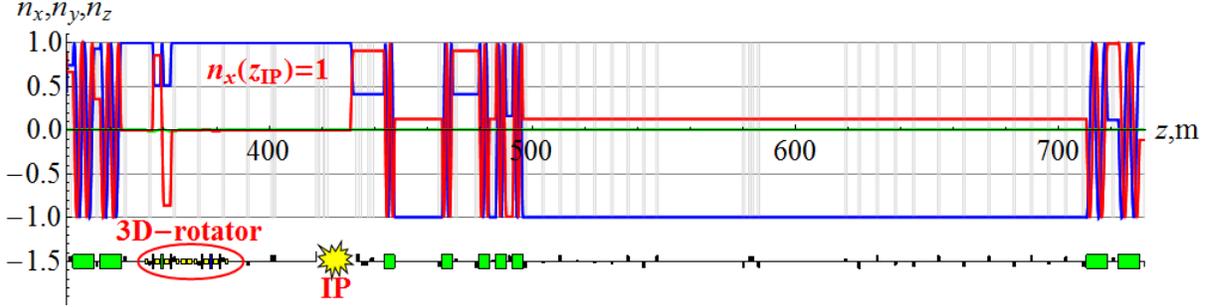

**Figure 16:** Proton beam's polarization in the experimental straight of the ion collider ring with $n_x(z_{IP}) = 1$.

The presented examples demonstrate the flexibility in control of the ion polarization when using a universal 3D spin rotator. The 3D spin rotator allows for a simple adjustment of the spin motion in the collider when modifying the magnetic lattice of the interaction region or introducing additional magnets into the lattice in the future. The only requirement on the introduced fields is that the accelerator remains transparent to the spin.

## 4. POLARIZATION STABILITY IN MEIC

### 4.1. SPIN RESONANCE STRENGTH AND POLARIZATION STABILITY CONDITION

The spin motion in MEIC is governed by the Thomas-BMT equation. In accelerator reference frame with basis unit vectors $\vec{e}_x$, $\vec{e}_y$, $\vec{e}_z$ tied to the beam design orbit (see Fig. 17), this equation has the form

$$\frac{d\vec{S}}{d\theta} = [\vec{W} \times \vec{S}], \quad \vec{W} = \gamma G \vec{K} + \vec{w}.$$

Here $\theta = 2\pi z/L$ is the generalized azimuthal angle (normalized distance along reference orbit), $\vec{K}$ is curvature of the design orbit.

The spin perturbation $\vec{w}$ contains two parts. The first part of the perturbation arises due to deviations of particles for the ideal design orbit. Orbit deviations are related to errors of the magnetic lattice as well as to beam emittances. The second part of the spin perturbation has to do with effect of control solenoids stabilizing the spin motion in MEIC.

Components of the spin perturbation $\vec{w}$ can be written in the accelerator reference frame. In the linear approximation ($\gamma G \gg 1$), they are equal to

$$w_x = -\gamma G \tau'_y, \quad w_y = \gamma G \tau'_x, \quad w_z = (1+G)h_z,$$

where vector $\vec{\tau} = (\tau_x, \tau_y, \tau_z)$ specifies the velocity direction of a beam particle, $\vec{\tau}' = d\vec{\tau}/d\theta$ is its derivative with respect to the generalized azimuthal angle, and $h_z = \frac{B_z L}{2\pi B \rho}$ is the normalized strength of the control solenoids setting the polarization direction $\vec{n}$ and the spin tune $\nu$.

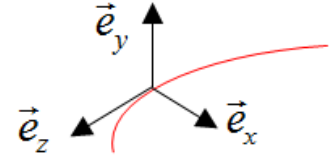

**Figure 17:** Basis unit vectors of the accelerator reference frame.

For stability of the polarization in the collider, the spin tune $\nu$ induced by the control solenoids must significantly exceed the strength of the zero-integer spin resonance:

$$\nu \gg w_0.$$

The resonance strength is determined by the spin perturbation component transverse to $\vec{n}$

$$w_0 = |\vec{w} - (\vec{w}\vec{n})\vec{n}|.$$

### 4.2. SPIN FIELD WITHOUT CONTROL SOLENOIDS IN MEIC

4.2.1 *Spin field components in the spin reference frame.*

Let us find spin field components for the case when the control solenoids in the 3D rotator are turned off. This problem can be solved in the accelerator reference frame. However, the spin field components have simpler form in the spin reference frame.

The basis unit vectors of the spin reference frame rotate in the collider arcs together with the spin of a particle moving along the design orbit. Therefore, the spin components of such a particle remain constant in the spin reference frame.

The relation between the basis unit vectors of the accelerator and spin reference frames is given by the following expressions:

$$\vec{e}_1 + i\vec{e}_3 = (\vec{e}_x + i\vec{e}_z) e^{-i\Psi_y}, \quad \vec{e}_2 = \vec{e}_y.$$

where $\Psi_y = \gamma G \int_0^\theta K_y d\theta$ is the spin rotation angle in the collider's bending dipoles.

In the linear approximation, averaging the spin perturbation gives the following spin field components in the spin reference frame:

$$\begin{cases} \omega_1 + i\omega_3 = -\gamma G \langle \tau'_y e^{i\Psi_y} \rangle = \gamma G \langle \tau_y \frac{de^{i\Psi_y}}{d\theta} \rangle \\ \omega_2 = \gamma G \langle \tau'_x \rangle = 0 \end{cases}$$

where brackets $\langle ... \rangle$ denote averaging over the particle's azimuthal angle. The vertical component of the spin field in the linear approximation turns into zero. The spin field vector lies in the orbital plane and is determined primarily by vertical excursions of particle orbits.

4.2.2 *Coherent part of the spin field*

The most substantial contribution to the spin field in colliders transparent to the spin comes from imperfections of a real magnetic lattice introducing radial perturbing fields. Such fields arise, for example, due to dipole roll errors, vertical quadrupole misalignments, etc. They result in vertical closed orbit distortion, are periodic and determine the coherent part of the spin field. The spin field induced by perturbing radial field $h_x(\theta)$, can be calculated using a periodic response function $F(\theta)$ [17]:

$$\omega_1 + i\omega_3 = \gamma G \langle h_x(\theta) F(\theta) \rangle.$$

For a flat figure-8 design orbit, the response function is expressed through the Floke function of vertical betatron oscillations $f_y(\theta)$ possessing the property $f_y(\theta + 2\pi) = e^{2\pi i \nu_b} f_y(\theta)$:

$$F(\theta) = \frac{f_y^*}{2} \int_{-\infty}^{\theta} \left(\frac{de^{-i\Psi_y}}{d\theta}\right) \frac{df_y}{d\theta} d\theta - \frac{f_y}{2} \int_{-\infty}^{\theta} \left(\frac{de^{-i\Psi_y}}{d\theta}\right) \frac{df_y^*}{d\theta} d\theta.$$

Figures 18 and 19 show graphs of the response function in the ion collider ring of MEIC for proton and deuteron beams, respectively, at 100 GeV/$c$.

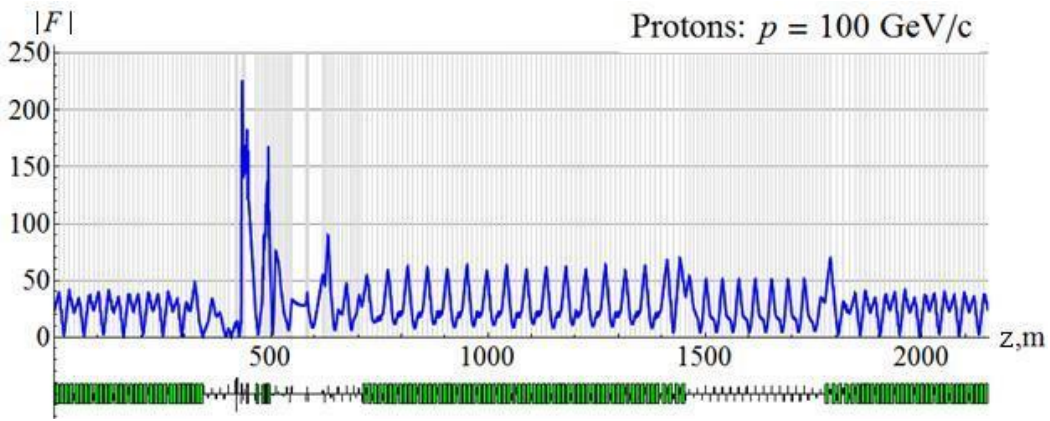

**Figure 18:** Response function for a proton beam at 100 GeV/$c$.

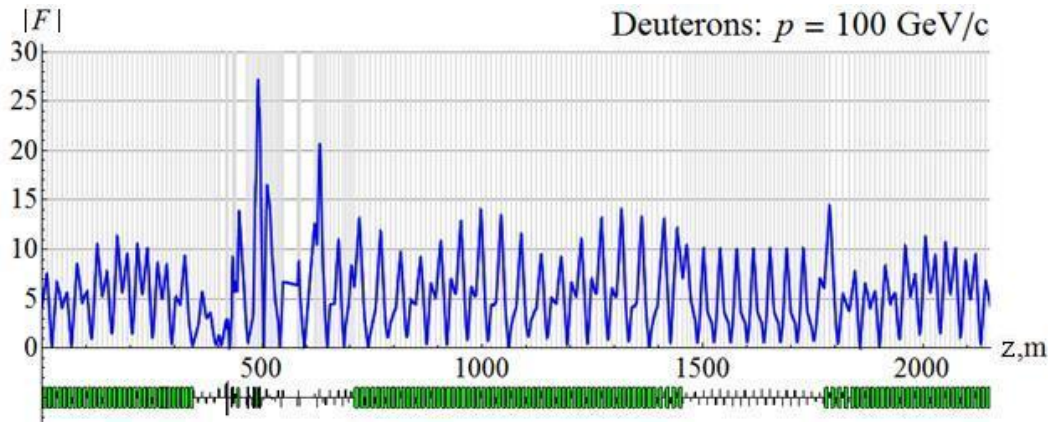

**Figure 19:** Response function for a deuteron beam at 100 GeV/$c$.

Let us calculate the spin field using a statistical model of quadrupole misalignments in vertical direction. Figures 20 and 21 show calculated dependence of the coherent part of the spin resonance strength $\omega_{\text{coherent}}$ on the proton and deuteron momenta, respectively, when the orbit excursion in the arcs does not exceed ±100 μm. A graph of an rms closed orbit distortion in the collider under the same parameters is shown in Fig. 22.

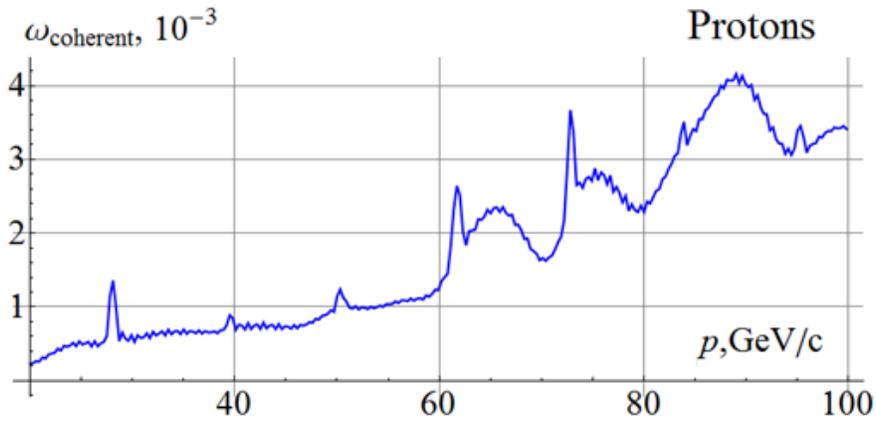

**Figure 20:** Coherent part of the spin field for a proton beam with random misalignment of all quadrupoles in the MEIC ion collider ring. The orbit excursion in the arcs does not exceed ±100 μm.

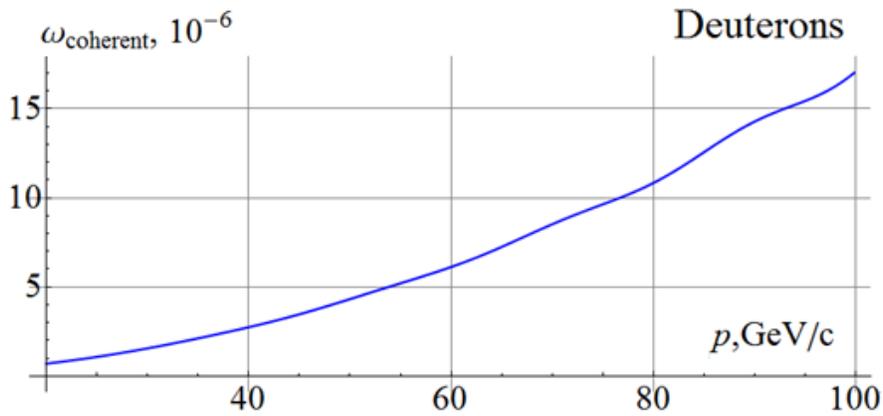

**Figure 21:** Coherent part of the spin field for a deuteron beam with random misalignment of all quadrupoles in the MEIC ion collider ring. The orbit excursion in the arcs does not exceed ±100 μm.

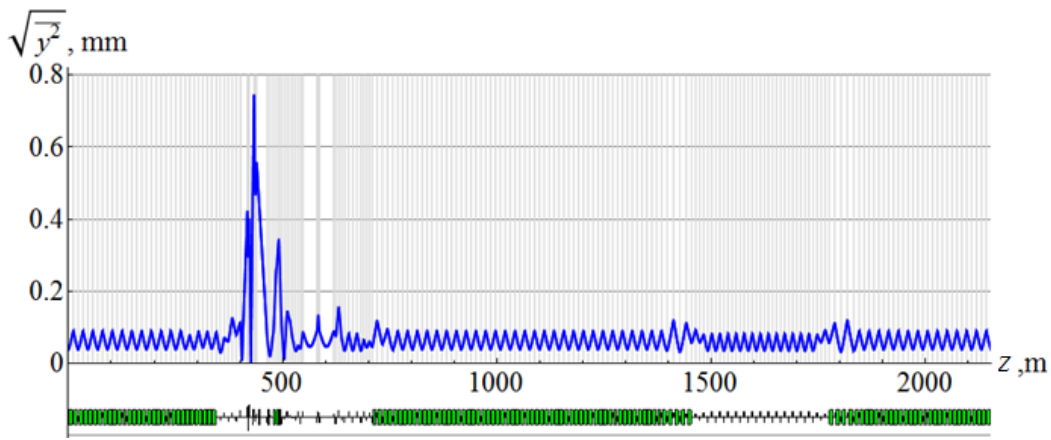

**Figure 22:** rms vertical beam excursion with random misalignment of all quadrupoles in the ion collider ring.

The greatest contribution to the spin resonance strength and rms orbit excursion comes from misalignment of strong quadrupoles near the interaction point. In the considered model, with the same distributions of alignment errors for all quadrupoles, contribution of these quadrupoles is an order of magnitude greater than that of the arc quadrupoles. Magnetic lattice in the experimental straight requires careful implementation and alignment. In principle, one can eliminate the effect of the strong quadrupoles in the experimental straight on the coherent part of the spin field. All that is needed is to make the response function zero near the interaction point by an appropriate choice of lattice structure.

4.2.3 *Compensation of the coherent part of the spin field*

The coherent part of the spin field is determined by three parameters and can be compensated by a 3D spin rotator with constant fields. Schematic of 3D rotators' placement in the MEIC ion collider ring is shown in Fig. 23.

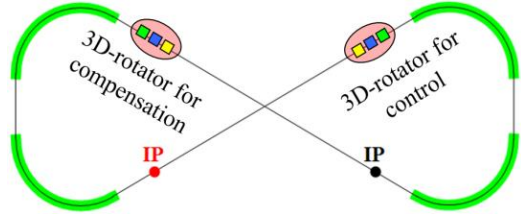

**Figure 23:** Schematic of 3D rotators' placement in the MEIC ion collider ring.

The first 3D rotator is located in the straight containing the interaction point and directly controls the polarization. The second 3D rotator with constant solenoid fields is located in the other straight and is used to compensate the coherent part of the zero-integer spin resonance strength. This allows one to significantly improve the polarized beam parameters as well as to greatly reduce the field integrals of the solenoids used for polarization control in the first rotator. In particular, the spin reversal time of the spin-flipping system of the MEIC ion collider ring will be on the order of 1 ms instead of 1 s.

The technique for compensation of integer resonance harmonics is well known and has been successfully utilized, for example, at the AGS [18].

4.2.4 *The incoherent part of the spin field*

In the linear approximation, the incoherent part of the spin field is related to the synchrotron modulation of energy caused by the accelerator's RF system:

$$\omega_1 + i\omega_3 = G\Delta\gamma \left\langle \frac{dD_y}{d\theta} \frac{de^{i\Psi_y}}{d\theta} \right\rangle,$$

where $D_y$ is the vertical dispersion in dipole magnets. At a high enough synchrotron tune ($\nu_{\text{sinch}} \gg \omega$), contribution of the synchrotron modulation becomes small due to averaging over time. Synchrotron modulation at a low frequency ($\nu_{\text{sinch}} \ll \omega$) splits a single zero-integer resonance into a series of satellite resonances separated from each other by the synchrotron modulation frequency. The greatest danger comes from the satellite resonances located in the range of $\sim \omega$, which must be avoided to preserve the polarization [17, 19].

The depolarizing effect of the synchrotron modulation is completely eliminated in the linear approximation in the case when the accelerator's magnetic lattice satisfies the condition

$$\left\langle \frac{dD_y}{d\theta} \frac{de^{i\Psi_y}}{d\theta} \right\rangle = 0.$$

In an ideal MEIC lattice, there is no vertical dispersion $D_y$ and this condition is automatically satisfied. Thus, in a figure-8 collider with conventional arcs (without radial-field magnets), the incoherent part of the resonance strength is calculated in the next-order approximation. In the second-order approximation, the spin field is directed vertically and is given by

$$\omega = \frac{\gamma^2 G^2}{2} \langle \tau_y^2 \frac{d\Psi_y}{d\theta} + \mathrm{Im}\, \tau_y \left(\frac{de^{-i\Psi_y}}{d\theta}\right) \int_{-\infty}^{\theta} \tau_y \left(\frac{de^{i\Psi_y}}{d\theta}\right) d\theta \rangle .$$

The spin perturbation $w_x$ in the arcs is determined by vertical betatron oscillations and can be written through the vertical Floke function $f(\theta)$. Using $\tau_y = C_y f_y' + C_y^* f_y'^*$ and expressing vertical betatron oscillation amplitude $C_y$ through the beam emittance $\left(|C_y|^2 = \varepsilon_y/4\right)$, we get a formula for the incoherent part of the spin field:

$$\omega = \frac{\varepsilon_y \gamma^2 G^2}{8\pi} \left[ \int_0^{2\pi} d\xi\, \frac{d\Psi_y(\xi)}{d\xi} |f_y'(\xi)|^2 + \right.$$

$$\left. + \frac{1}{2} \mathrm{Im} \int_0^{2\pi} d\xi \int_{-\infty}^{\xi} d\zeta \left(\frac{de^{-i\Psi_y(\xi)}}{d\xi}\right)\left(\frac{de^{-i\Psi_y(\zeta)}}{d\zeta}\right) \left(f_y'(\xi) f_y'^*(\zeta) + f_y'^*(\xi) f_y'(\zeta)\right) \right].$$

Figures 24 and 25 show graphs of the incoherent parts of the proton and deuteron resonance strengths $\omega_{\text{incoherent}}$ versus momentum for a normalized beam emittance $\varepsilon_0$ of 0.07 μm·rad. The incoherent part of the spin resonance strength is related to asymmetry in the collider's magnetic lattice.

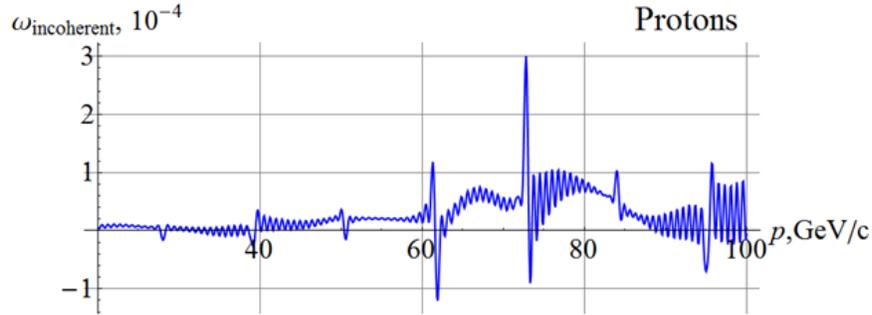

**Figure 24:** Incoherent part of the resonance strength for a proton beam.

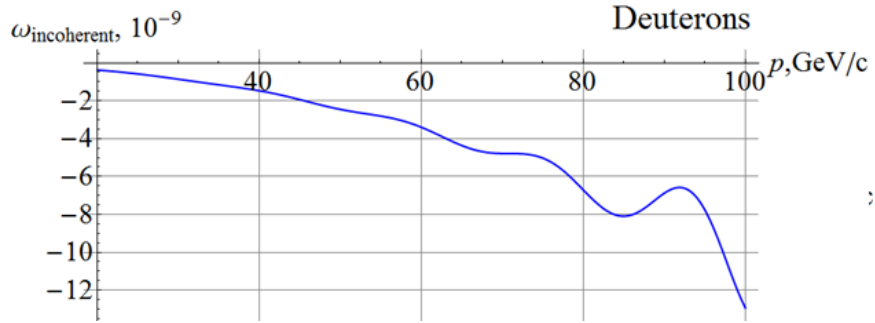

**Figure 25:** Incoherent part of the resonance strength for a deuteron beam.

A numerical calculation shows that, in the MEIC ion collider lattice, contribution of the incoherent part of the spin field is a few orders of magnitude lower than that of the coherent part.

## 5. CONCLUSIONS

Let us briefly summarize the main features of the figure-8 design. Such a design allows for:
- use of weak solenoids for polarization control at high energies of any particle species including deuterons;
- seamless integration of the spin control elements into the collider lattice with fixed closed orbit and no optics distortion;
- elimination of the resonant depolarization at all stages of the beam acceleration from the linac to the collider ring;
- adjustment of any polarization orientation at any orbital location (spin matching at injection into the different accelerator complex components, polarimetry, spin flipping);
- manipulation of the particle spin during an experiment without affecting the beam orbital properties, which provides a capability of carrying out polarized beam experiments at a new precision level;
- compensation of manufacturing and alignment errors of the lattice magnetic elements, which additionally substantially enhances the precision of polarized beam experiments;
- ease of adjusting the spin dynamics to meet any experimental requirements, which may arise in the future.

The preliminary numerical analysis of proton and deuteron polarization stability in the ion collider ring of MEIC allows one to draw the following conclusions:
- The polarization control insertion does not affect the orbital beam parameters of the MEIC ion collider ring.
- For stability of ion polarization in MEIC, the spin tune induced by the 3D spin rotator must significantly exceed the strength of the zero-integer spin resonance.
- Calculations of the resonance strength for MEIC show that its coherent part related to closed orbit distortion is a few orders of magnitude greater than its incoherent part related to beam emittances.
- 3D rotators with 2 T solenoids provide control of proton and deuteron polarizations in MEIC.
- Polarized beam quality can be additionally significantly improved and the field strengths of the control solenoids in the 3D rotator can be significantly reduced by compensating the coherent part of the resonance strength.

The results of this work have been presented at SPIN'14 [7], IPAC'15 [8, 16, 20], and discussed at an accelerator seminar at Jefferson Lab on May 14, 2015. The obtained results were regularly discussed at teleconferences with JLab's CASA staff.

## ACKNOWLEDGEMENTS

This work was supported by Jefferson Science Associates, LLC under U.S. DOE Contracts No. DE-AC05-06OR23177 and DE-AC02-06CH11357. The U.S. Government retains a non-exclusive, paid-up, irrevocable, world-wide license to publish or reproduce this manuscript for U.S. Government purposes.